\documentclass[5p,times]{elsarticle}

\journal{Physics Letters B}

\usepackage{amssymb}
\def\MeV{\,MeV}
\def\GeV{\,GeV}
\def\kevc1{\ifmmode\mathrm{\ keV/{\mit c}}
          \else$\mathrm{\ keV/{\mit c}}$\fi}

\def\MeVc1{\,MeV/{\mit c}}

\def\mevc1{\ifmmode\mathrm{\ MeV/{\mit c}}
          \else$\mathrm{\ MeV/{\mit c}}$\fi}
\def\gevc1{\ifmmode\mathrm{\ GeV/{\mit c}}
          \else$\mathrm{\ GeV/{\mit c}}$\fi}
\def\GeVc1{\ifmmode\mathrm{\ GeV/{\mit c}}
          \else$\mathrm{\ GeV/{\mit c}}$\fi}
\def\kevc2{\ifmmode\mathrm{\ keV/{\mit c}^2}
          \else$\mathrm{\ keV/{\mit c}^2}$\fi}
\def\Mevc2{\ifmmode\mathrm{\ MeV/{\mit c}^2}
          \else$\mathrm{\ MeV/{\mit c}^2}$\fi}
\def\Gevc2{\ifmmode\mathrm{\ GeV/{\mit c}^2}
          \else$\mathrm{\ GeV/{\mit c}^2}$\fi}
\def\Gev2c2{\ifmmode\mathrm{\ GeV^2/{\mit c}^2}
          \else$\mathrm{\ GeV^2/{\mit c}^2}$\fi}
\def\Pgp{\ifmmode\math{p}
         \else$\mathrm{p}$\fi}
\def\Pagp{\ifmmode\mathrm{\overline{p}}
         \else$\mathrm{\overline{p}}$\fi}
\def\Pgn{\ifmmode\mathrm{n}
         \else$\mathrm{n}$\fi}
\def\Pagpn{\ifmmode\mathrm{\overline{n}}
         \else$\mathrm{\overline{n}}$\fi}
\def\Pp{\ifmmode\mathrm{p}
         \else$\mathrm{p}$\fi}
\def\Pap{\ifmmode\mathrm{\overline{p}}
         \else$\mathrm{\overline{p}}$\fi}

\def\Pn{\ifmmode\mathrm{n}
         \else$\mathrm{n}$\fi}
\def\Pan{\ifmmode\mathrm{\overline{n}}
         \else$\mathrm{\overline{p}}$\fi}
\def\Py{\ifmmode\mathrm{Y}
         \else$\mathrm{Y}$\fi}
\def\Pay{\ifmmode\mathrm{\overline{Y}}
         \else$\mathrm{\overline{Y}}$\fi}

\def\PgL{\ifmmode\mathrm{\Lambda}
          \else$\mathrm{\Lambda}$\fi}
\def\PagL{\ifmmode\mathrm{\overline{\Lambda}}
            \else$\mathrm{\overline{\Lambda}}$\fi}
\def\PgS{\ifmmode\mathrm{\Sigma}
          \else$\mathrm{\Sigma}$\fi}
\def\PagS{\ifmmode\mathrm{\overline{\Sigma}}
            \else$\mathrm{\overline{\Sigma}}$\fi}
\def\PgSp{\ifmmode\mathrm{\Sigma^+}
          \else$\mathrm{\Sigma^+}$\fi}
\def\PagSp{\ifmmode\mathrm{\overline{\Sigma^+}}
            \else$\mathrm{\overline{\Sigma^+}}$\fi}
\def\PgSm{\ifmmode\mathrm{\Sigma^-}
          \else$\mathrm{\Sigma^-}$\fi}
\def\PagSm{\ifmmode\mathrm{\overline{\Sigma^-}}
            \else$\mathrm{\overline{\Sigma^-}}$\fi}
\def\PgSn{\ifmmode\mathrm{{\Sigma}^0}
            \else$\mathrm{{\Sigma}^0}$\fi}
\def\PagSn{\ifmmode\mathrm{\overline{\Sigma}^0}
            \else$\mathrm{\overline{\Sigma}^0}$\fi}
\def\PgX{\ifmmode\mathrm{\Xi}
          \else$\mathrm{\Xi}$\fi}
\def\PagX{\ifmmode\mathrm{\overline{\Xi}}
            \else$\mathrm{\overline{\Xi}}$\fi}
\def\PgXm{\ifmmode\mathrm{\Xi^-}
          \else$\mathrm{\Xi^-}$\fi}
\def\PagXm{\ifmmode\mathrm{\overline{\Xi^-}}
            \else$\mathrm{\overline{\Xi^-}}$\fi}
\def\PagXp{\ifmmode\mathrm{\overline{\Xi}^+}
            \else$\mathrm{\overline{\Xi}^+}$\fi}

\def\PgOm{\ifmmode\mathrm{\Omega^-}
          \else$\mathrm{\Omega^-}$\fi}
\def\PagOm{\ifmmode\mathrm{\overline{\Omega^-}}
            \else$\mathrm{\overline{\Omega^-}}$\fi}
\def\PgOp{\ifmmode\mathrm{\Omega^+}
          \else$\mathrm{\Omega^+}$\fi}
\def\PagOp{\ifmmode\mathrm{\overline{\Omega}^+}
            \else$\mathrm{\overline{\Omega}^+}$\fi}

\def\PgLc{\ifmmode\mathrm{\Lambda_c}
          \else$\mathrm{\Lambda_c}$\fi}
\def\PagLc{\ifmmode\mathrm{\overline{\Lambda}_c}
            \else$\mathrm{\overline{\Lambda}_c}$\fi}

\def\PgD{\ifmmode\mathrm{D}
          \else$\mathrm{D}$\fi}
\def\PagD{\ifmmode\mathrm{\overline{D}}
            \else$\mathrm{\overline{D}}$\fi}

\def\PgPi{\ifmmode\mathrm{\pi}
          \else$\mathrm{\pi}$\fi}
\def\PagPi{\ifmmode\mathrm{\overline{\pi}}
            \else$\mathrm{\overline{\pi}}$\fi}

\begin{document}

\begin{frontmatter}



%% e.g. \dochead{17th International Conference on Dynamical Processes in Excited States of Solids}

\title{Antihyperon potentials in nuclei via exclusive antiproton-nucleus reactions at FAIR}

\author[a]{Alicia Sanchez Lorente}
\author[a]{Sebastian Bleser}
\author[a]{Sebastian Bleser}
\author[a,b]{Josef Pochodzalla\corref{correspondingauthor}   }
\cortext[correspondingauthor]{Corresponding author}
\ead{pochodza@kph.uni-mainz.de}
\address[a]{Helmholtz Institute Mainz, Johannes Gutenberg-Universit\"at Mainz, D-55099 Mainz, Germany}
\address[b]{Institut f\"ur Kernphysik, Johannes Gutenberg-Universit\"at Mainz, D-55099 Mainz, Germany}

\begin{abstract}
The exclusive production of hyperon-antihyperon pairs close to their production threshold in $\Pap$ - nucleus collisions
offers a unique and hitherto unexplored opportunity to elucidate the behaviour of antihyperons in nuclei. For the first time we analyse
these reactions in a microscopic transport model using the  the Gie\ss en Boltzmann-Uehling-Uhlenbeck transport model.
The calculation take the delicate interplay between the strong absorption of antihyperons, their rescattering and refraction
at the nuclear surface as well as the Fermi motion of the struck nucleon into account.
We find a substantial sensitivity of transverse momentum correlations of coincident $\PagL\PgL$-pairs to the assumed depth of the $\PagL$-potential.
Because of the high cross section for this process and the simplicity of the
experimental method our results are highly relevant for future activities at the international Facility
for Antiproton and Ion Research (FAIR).
\end{abstract}


\end{frontmatter}

\section{Introduction}
\label{sec01}
The interaction of individual baryons or antibaryons in nuclei provides a unique opportunity to elucidate strong in-medium effects in baryonic systems. While for neutrons and protons as well as some strange baryons experimental information on their binding in nuclei exists, information on antibaryons in nuclei are rather scarce. Only for the antiproton the nuclear potential could be constrained by experimental studies. The (Schr\"odinger equivalent) antiproton potential at normal nuclear density turns out to be in the range of $U_{\overline{p}} \simeq -150$MeV, i.e. a factor of approximately 4 weaker than expected from naive G-parity relations \cite{Lar09}. Gaitanos {\em et al.} \cite{Gai11} suggested that this discrepancy can be traced back to the missing energy dependence of the proton-nucleus optical potential in conventional relativistic mean-field models. The required energy and momentum dependence could be recovered by extending the relativistic hadrodynamics Lagrangian by non-linear derivative interactions  \cite{Gai09,Gai11,Gai13} thus also mimicking many-body forces \cite{Gom14}. Considering the important role played e.g. by strange baryons and antibaryons production for a quantitative interpretation of high-energy heavy-ion collisions and dense hadronic systems it is clearly mandatory to test these concepts also in the strangeness sector. Furthermore, it was pointed out recently \cite{Lar10} that in-medium interactions of antibaryons may cause compressional effects and may thus provide additional information on the nuclear equation-of-state \cite{Gai15}. Therefore, the question if and to what extend G-parity is violated by antihyperons in nuclei is also a challenging problem by itself.

Antihyperons annihilate quickly in nuclei and conventional spectroscopic studies of bound systems comparable
to hypernuclei are not feasible. As a consequence, no experimental information on the nuclear potential of antihyperons exists so far.
As suggested recently \cite{Poc08}, quantitative information on the antihyperon potentials may be
obtained via exclusive antihyperon-hyperon pair production close to threshold in antiproton-nucleus interactions.
Once these hyperons and antihyperons leave the nuclear environment they can be detected and their asymptotic
momentum distributions will reflect the depth of the respective
potentials. In ref. \cite{Poc08} it was demonstrated that momentum correlations of emitted hyperon-antihyperon
pairs can be used to extract information on the relative potential of hyperons and antihyperons in nuclei.
Since in the $\Pap\Pp$ center-of-mass the distribution of the free baryon-antibaryon pairs is non-isotropic, the analy\-sis relied mainly on the {\em transverse} momenta of the coincident baryons and antibaryons. The calculations of Ref.
\cite{Poc08} revealed significant sensitivities of the transverse momentum
asymmetry $\alpha_{T}$ which is defined in terms of the
transverse momenta of the coincident particles
\begin{equation}
\alpha_{T}=\frac{p_{T}(\PgL)-p_{T}(\PagL)}{p_{T}(\PgL)+p_{T}(\PagL)}.
\label{eq:01}
\end{equation}
to the depth of the antihyperon potential. The asymmetry
 $\alpha_{T}$ turned out to be rather robust in case of model
parameter variations and remained substantial also if
the momentum dependence of the potential was considered
\cite{Poc09}. However, these sche\-matic simulations ignored
rescattering processes and refractive effects at the potential
boundary. These effects are expected to erode
the two-body character of the $\PagL\PgL$  production and may
thus diminish or even destroy the sensitivity. In order to
go beyond the schematic calculations presented in Refs.
\cite{Poc08,Poc09} and to include simultaneously rescattering, refraction
and absorption effects, we present here first realistic
calculations of this new observable with a microscopic
transport model.

\begin{figure}[tb]
\includegraphics[width=0.49\textwidth]{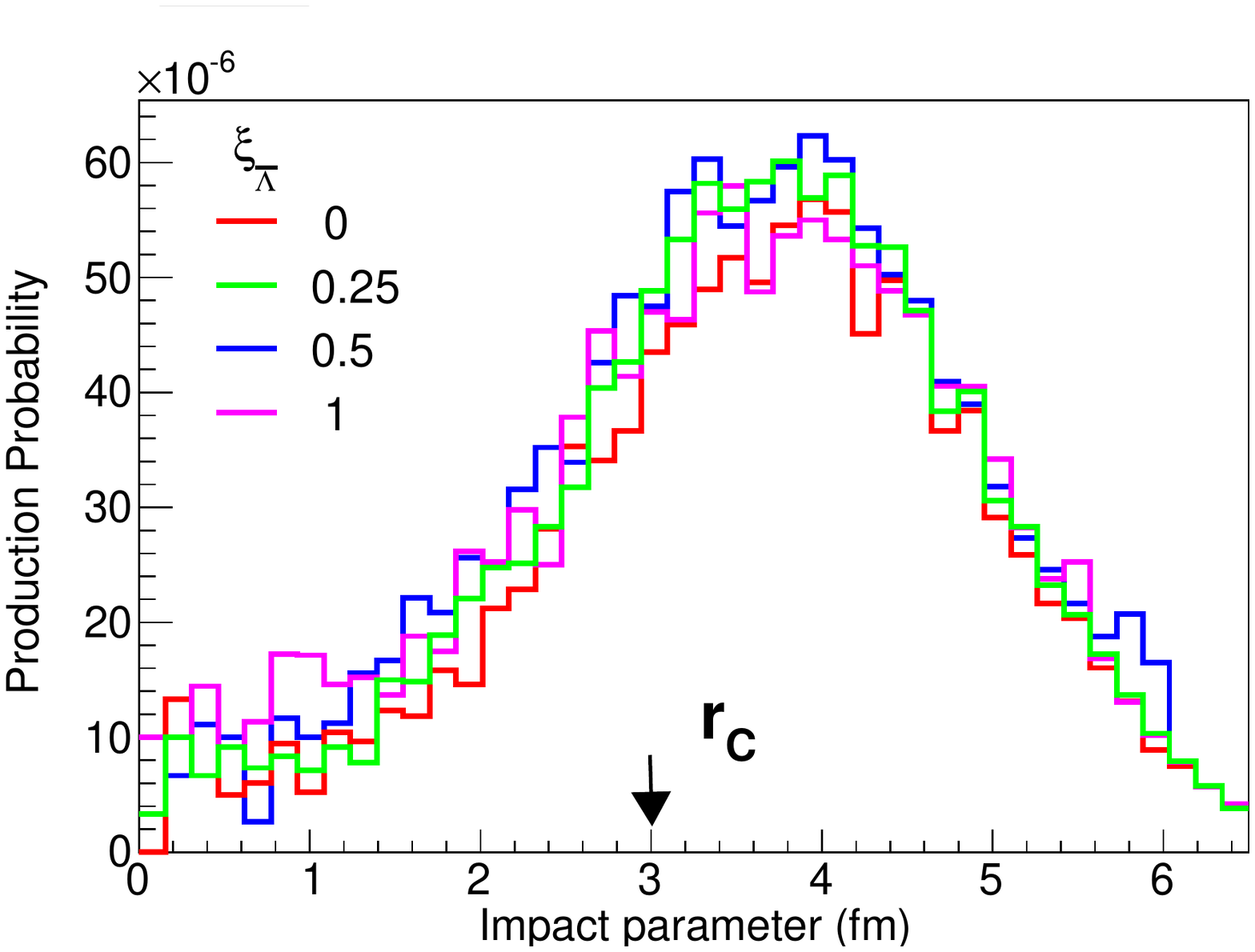}
\includegraphics[width=0.49\textwidth]{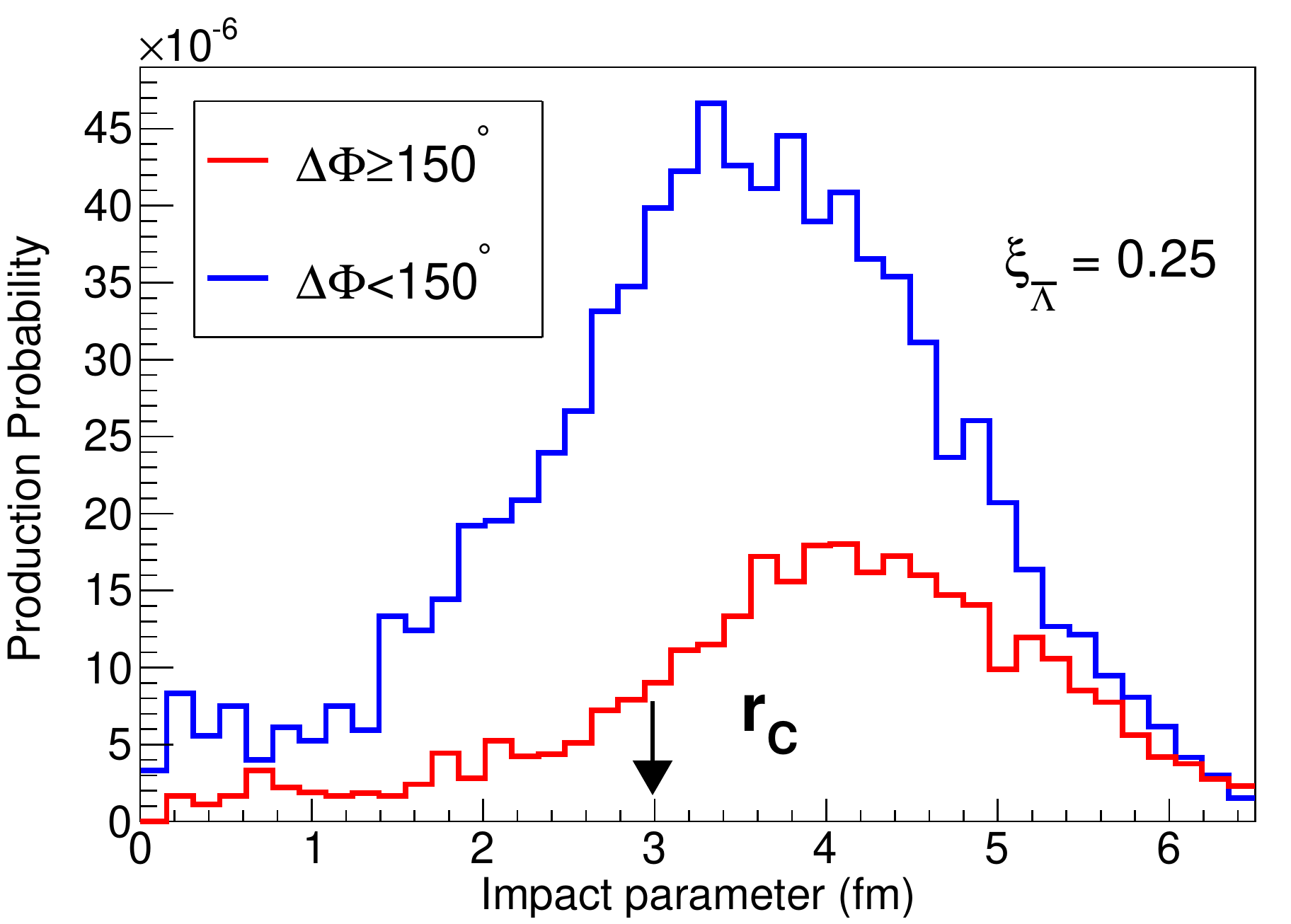}
\caption{Top: Probability distribution for free {\PgL\PagL}-pair production in 0.85\GeV\
\Pap +$^{20}$Ne collisions as a function of the impact parameter. The different lines show the GiBUU predictions for different scaling factor
$\xi_{\PagL}$ of the $\PagL$-potentials. Bottom: Impact parameter distributions for planar ($\Delta\Phi \geq 150^o$) and non-planar ($\Delta\Phi < 150^o$) {\PgL\PagL} pairs (cf. Fig.~\ref{fig:02}) using a fixed $\PagL$-potential scaling factor $\xi_{\PagL}$ = 0.25.
The arrows mark the rms-charge radius r$§_C$ of $^{20}$Ne}
\label{fig:01}
\end{figure}

\section{Transport calculations of antihyperon-hyperon production}
\label{sec02}

The Giessen Boltzmann-Uehling-Uhlenbeck transport modell (GiBUU, Release 1.5) \cite{GiBUU} describes many features of \Pagp-nucleus interactions in the FAIR energy range \cite{Lar09,GiBUU,Lar12}. Particularly the presently available data on strangeness production are well reproduced. In this code non-linear derivative interactions \cite{Gai11} are not yet included and a simple scaling factor $\xi_{\Pagp}$\,=\,0.22 is applied to ensure a Schr\"odinger equivalent antiproton potential of $-$150\MeV\ at saturation density \cite{Lar12}. (Note that this value differs slightly from the scaling factor $\xi_{\Pagp}$\,=\,0.25 given in Ref. \cite{Lar09}.) The hyperon potentials were fixed by hypernuclear and hyperatom data \cite{Lar12}. No experimental information exists for antihyperons in nuclei and G-parity symmetry is therefore used to specify their default potentials.
This leads to a $\PagL$-potential U(\PagL)=-449\MeV. As already stressed in ref. \cite{Lar12}, the attractive $\PgS$-potential and the weak attraction for kaons adopted in the GiBUU
model is not compatible with experimental data. In the
present work we focus on exclusive $\PagL\PgL$-pair production
in the threshold region. Therefore, we do not expect that
a more realistic treatment of the \PgS\ and kaon potentials
will modify the conclusions of the present study.

We have studied the exclusive reaction \Pap+$^{20}$Ne$\rightarrow$ \PagL\PgL\ at beam energies of 0.85\GeV\ and 1\GeV. These energies correspond to antiproton momenta of 1.522\GeVc1 and 1.696\GeVc1, respectively.
At 0.85\GeV\ the excess energy with respect to the elementary reaction \Pagp\ +\Pp$\rightarrow$\PagL\PgL\ amounts to only 30.6\MeV. Therefore, the \PagS\PgL\
and \PgS\PagL\ channels are not accessible and also the production of a pion in addition to a $\PgL\PagL$-pair can be neglected. The
higher energy of 1\GeV\ lies above the $\PagS\PgL$- threshold and
makes also the \Pap\Pn\ $\rightarrow\PagL\PgSm$ and \Pap\Pp\ $\rightarrow$\PagL\PgSn\ as well as their
charge conjugate channels accessible. Those channels will be discussed in a forthcoming paper.

\begin{figure}[bt]
\includegraphics[width=0.49\textwidth]{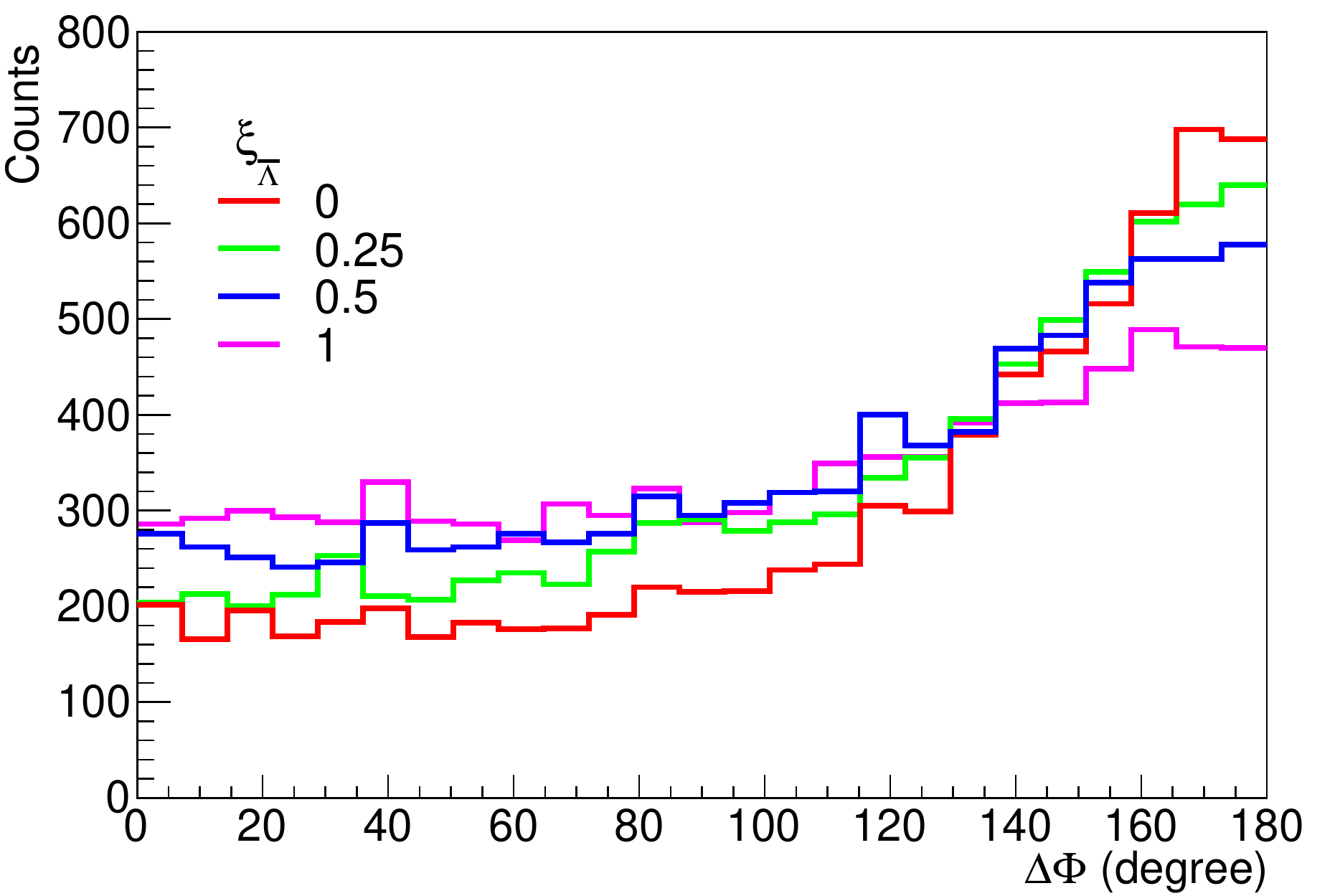}
\caption{Coplanarity of coincident $\PgL\PagL$-pairs produced exclusively in 0.85\GeV\
\Pap +$^{20}$Ne interactions. The different histograms show the GiBUU predictions for different scaling factor
$\xi_{\PagL}$ of the $\PagL$-potentials.}
\label{fig:02}
\end{figure}

In order to explore the sensitivity of the transverse momentum asymmetry on the depth of the $\PagL$-potential we
have performed a series of calculations where only the antihyperon potentials were modified by a single scaling
factor $\xi_{\PagL}$ , leaving all other input parameters of the model unchanged. The calculations were performed at
the High Power Computing Cluster HIMSTER located at the Helmholtz-Institute Mainz. Each GiBUU-Job comprised
1000 parallel events. In order to keep the necessary computing time low, all cross sections for antihyperon -- hyperon
pair production were artificially enhanced by a factor of 10. Since within the 1000 parallel events of an
individual job the probability of a multiple production of hyperon pairs is low, the mean field dynamics is insignificantly distorted. For each parameter set a total of
26460 Jobs were generated. Each parameter set shown
in the following contains approximately 8000 $\PagL\PgL$-pairs
where both, the $\PagL$ and the $\PgL$ escaped the nucleus.

Unlike in inclusive reactions \cite{Sibirtsev99,Lenske05}, the strong absorption
of the antihyperons in nuclei favors the production of
free hyperon-antihyperon pairs in the corona of the target
nucleus. Configurations where the path of the antihyperon within the target nucleus
is minimal or where secondary scattering reduces this path length  
towards the nuclear surface will be more likely to produce free $\PagL\PgL$-pairs.
The top part of Fig.~\ref{fig:01} shows the probability for
free $\PagL\PgL$-pair production ($\sim$b$^{-1}dN_{\PagL\PgL}/db$) as a function of
the impact parameter b for different $\PagL$-potentials. (The
artificial increase for antihyperon--hyperon production by a factor of 10
is taken into account.) The distributions peak around
3.8\,fm which is significantly larger than the $^{20}$Ne rms-charge
radius of r$§_C$ = 3.0\,fm \cite{DeV87} marked by the arrow in
Fig.~\ref{fig:01}. Consequently, free antihyperon-hyperon pairs are
mainly produced at low nuclear densities corresponding
to 20 to 25\% of the central density.

\section{Results and discussion}
\label{sec03}

For scaling factors $\xi_{\PagL}$ between 1 and 0.25 the average
impact parameter persists at 3.8\,fm. Only for $\xi_{\PagL}$=0 a
slight increase to 3.9\,fm is observed. At the same time the
number of produced $\PagL\PgL$-pairs remains fairly constant on
a level of about 8700 events over the range $1 \leq \xi_{\PagL}\leq 0.5$
and decreases for $\xi_{\PagL}$ = 0 by about 15\% to $\approx$ 7500 events.
This moderate decrease indicates that absorption does
not change dramatically with the depth of the $\PagL$-potential.
In Fig.~\ref{fig:01} (top) the main variations are seen at
small impact parameter. This suggests that the smaller
number of events in case of a shallow antihyperon potential
is caused by an increased absorption at more central
collisions. For a quantitative interpretation one has to
keep in mind, however, that also a potential-dependent
rescattering can enhance the escape probability from the
nucleus by decreasing the path length until the nuclear
surface of particularly the antihyperons.

The role of secondary deflection can be explored by
the coplanarity of the $\PagL\PgL$-pairs. (Of course, in these reactions
close to threshold the Fermi motion of the struck
proton inside the nuclear target contributes also to the
final momenta.) Fig.~\ref{fig:02} shows the angle $\Delta\Phi$ defined as the difference between the
azimuthal angles of the free $\PagL$ and \PgL. Already for zero
$\PagL$-potential, the coplanarity is strongly blurred. With
increasing potential depth for the $\PagL$, the coplanarity is
even less pronounced. The significant deviation from
180$^o$ demonstrates the influence of secondary scattering
prior to the emission of the $\PagL$ or $\PgL$ or a deflection at the
potential boundary.
\begin{figure}[bth]
\includegraphics[width=0.44\textwidth]{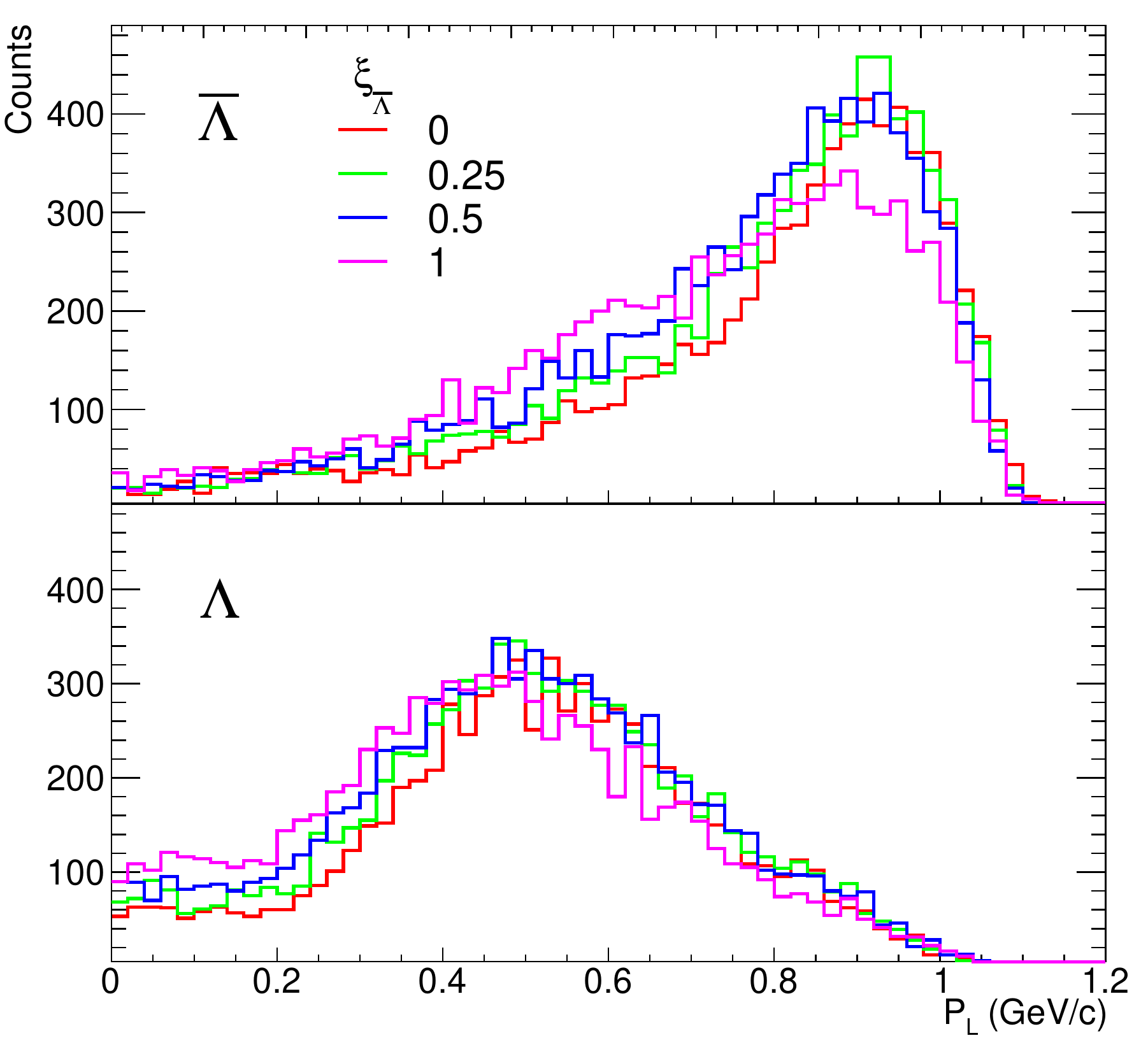}
\includegraphics[width=0.44\textwidth]{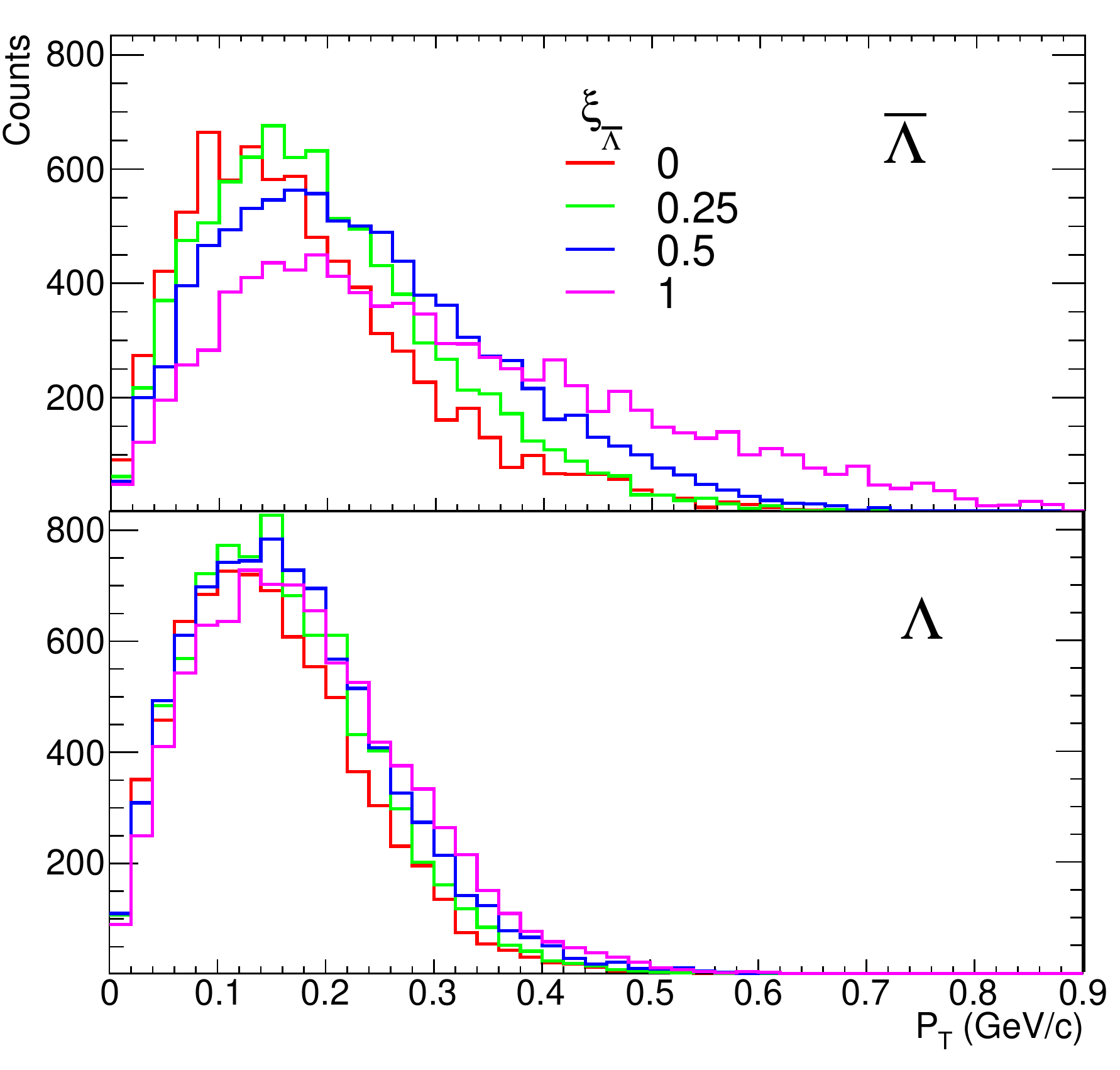}
\caption{Longitudinal (top) and transverse momentum distributions (bottom) for $\PgL$ and $\PagL$ from exclusive pair production. The different symbols show the GiBUU predictions for different scaling factors
$\xi_{\PagL}$ of the $\PagL$-potential.}
\label{fig:03}
\end{figure}

The effect of enhanced secondary scattering processes in the case of deep antihyperon potentials is also visible in the momentum distributions of the emitted particles.  Elastic scattering will on the average decrease the
longitudinal momentum and will broaden the transverse
momentum distribution. Fig.~\ref{fig:03} shows the longitudinal
momenta (top) and transverse momenta (bottom) for $\PagL$
and $\PgL$-hyperons from exclusive pair production. Since
the antihyperons are emitted more forward in the center
of mass, their (longitudinal) momentum distributions
peak at values around 0.9$\GeVc1$ , while for $\PgL$-hyperons
the typical momenta reach only half of this value. The decrease
of the longitudinal momenta for both, $\PgL$-hyperons
as well as $\PagL$-hyperons with deeper antihyperon potentials
and hence more rescattering (c.f. Fig.~\ref{fig:02}) is clearly seen
in the top part of Fig.~\ref{fig:03}. Because of the larger initial
longitudinal momenta, scattering in the target nucleus
is expected to lead to a larger absolute broadening of
the $\PagL$ transverse momentum distributions as compared
to the lower momentum $\PgL$-hyperons. As can be seen in
the lower part of Fig.~\ref{fig:03}, the increase of the transverse
momenta with deeper $\PagL$-potential is indeed more pronounced
for $\PagL$-hyperons than for $\PgL$-hyperons.
\begin{figure}[tb]
\includegraphics[width=0.49\textwidth]{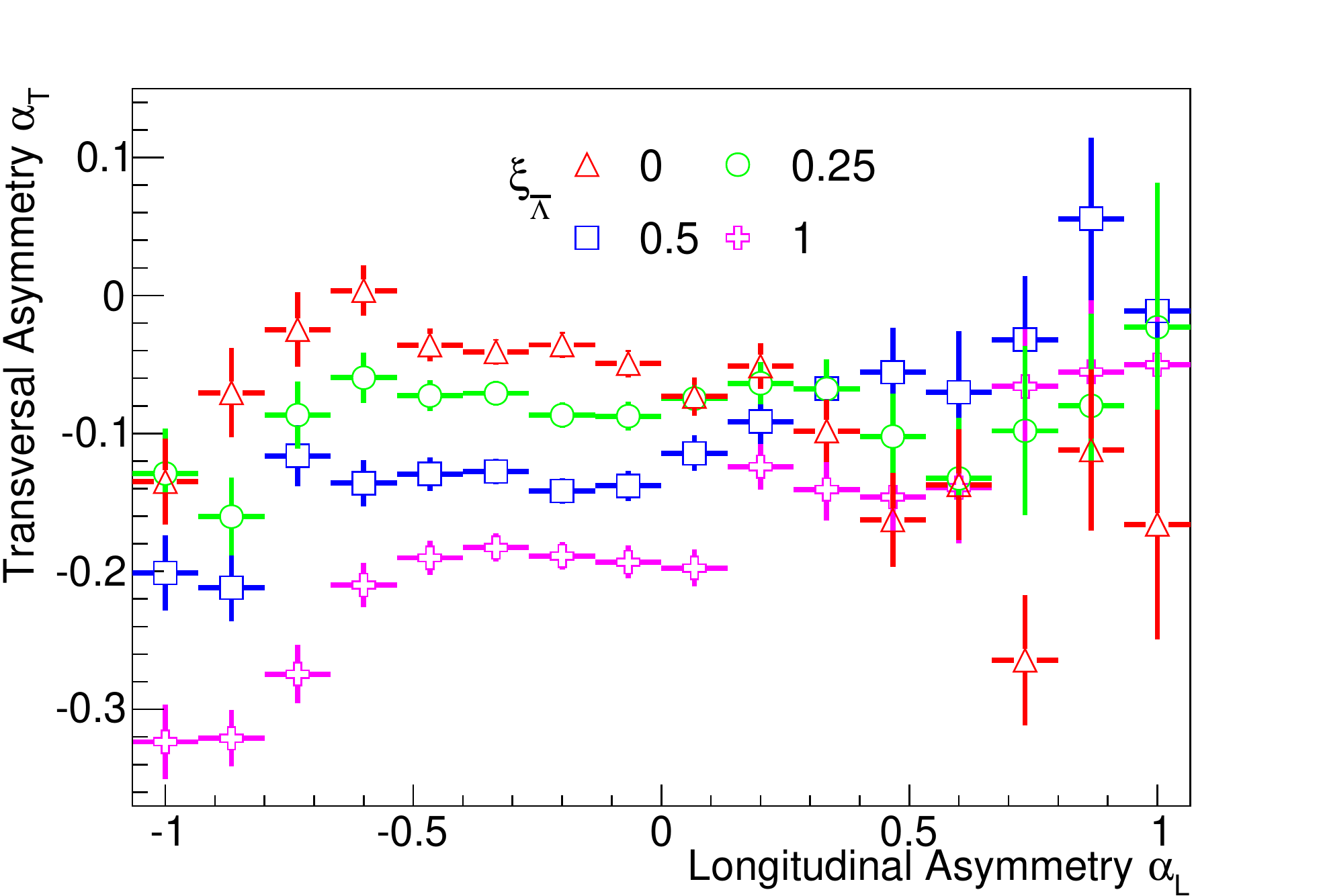}\\
\includegraphics[width=0.49\textwidth]{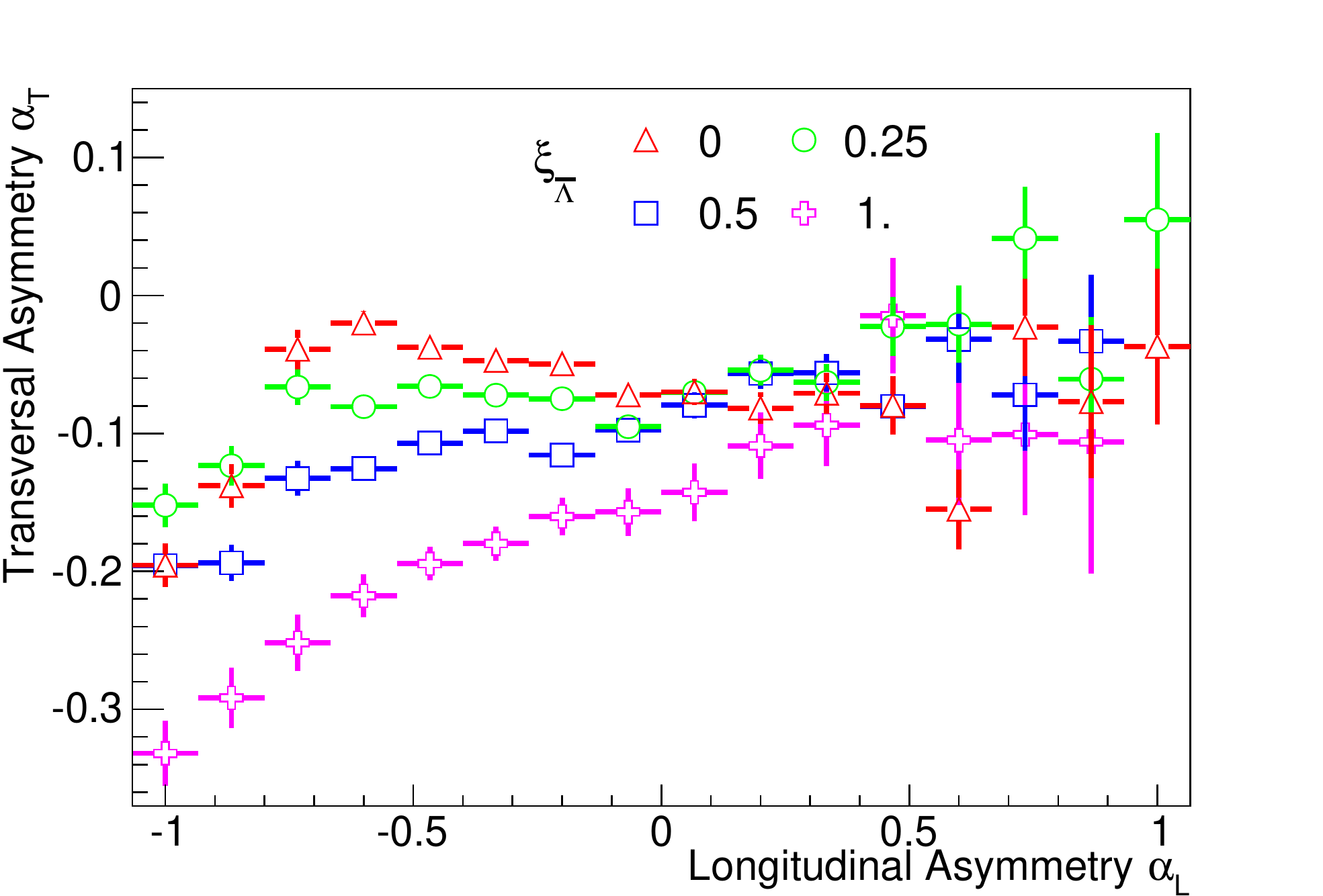}
\caption{Average transverse momentum asymmetry  $\alpha_{T}$ (Eq. \ref{eq:01}) as a function of the
longitudinal momentum asymmetry for $\PgL\PagL$-pairs produced
exclusively in 0.85\GeV\ (top) and 1\GeV\ (bottom)
\Pap+$^{20}$Ne interactions. The different symbols show the GiBUU predictions for different scaling factor $\xi_{\PagL}$ of the $\PagL$-potentials.}
\label{fig:04}
\end{figure}

The delicate interplay between the Fermi motion of the struck nucleon, the absorption, rescattering and refraction
at the nuclear surface of the produced hyperons and antihyperons is further illustrated in the bottom part of Fig.~\ref{fig:01}.
Coplanar $\PagL\PgL$-pairs with $\Delta\Phi\geq150^o$ are associated
with larger impact parameters (${\langle}b\rangle_{\Delta\Phi\geq150^o}$ = 4.1\,fm) as compared to the case $\Delta\Phi<150^o$  where scattering processes contribute more (${\langle}b\rangle_{\Delta\Phi<150^o}$ = 3.7\,fm).
The figure suggests that at small impact parameters the
rescattering is indeed crucial for antihyperons being able
to escape from the nucleus.

In Fig.~\ref{fig:04} we finally show the GiBUU prediction for the
transverse asymmetry $\alpha_{T}$ (Eq. \ref{eq:01}) for different scaling
factors  $\xi_{\PagL}$  of the $\PagL$-potential. As in Ref. \cite{Poc08} we plot the
average $\alpha_{T}$ as a function of the longitudinal momentum
asymmetry $\alpha_{L}$ which is defined for each event as
\begin{equation}
\noindent
\alpha_{L}=\frac{p_{L}(\PgL)-p_{L}(\PagL)}{p_{L}(\PgL)+p_{L}(\PagL)}.
\label{eq:02}
\end{equation}
For 0.85\GeV\ (top) as well as 1\GeV\ (bottom) antiproton energy a remarkable sensitivity of $\alpha_{T}$ on the $\PagL$-potential is found at negative values of $\alpha_{L}$.
Despite the concern mentioned in the introduction, secondary effects do not
wipe out the dependence of $\alpha_{T}$ on the antihyperon potential.
Both, this significant larger sensitivity as compared
to the schematic calculation in Ref. \cite{Poc08} as well as
the shift of the average $\alpha_{T}$ towards more negative values
are linked to the substantial transverse momentum
broadening for the $\PagL$-hyperons by secondary scattering
(c.f. lower part of Fig.~\ref{fig:03}). For positive values of $\alpha_{L}$
where the antihyperon is emitted backward with respect
to $\PgL$-particle, the statistics in the present simulation is
too low to draw quantitative conclusions. But even in this
region of $\alpha_{L}$ a systematic variation of $\alpha_{T}$ with the antihyperon
potential might show up with improved statistics.

The international Facility for Antiproton and Ion Research
(FAIR) presently under construction in Darmstadt
(Germany) will provide high intensity antiproton
beams with momenta between 1.5\GeVc1 and
15$\GeVc1$. A unique feature of antiproton interactions
in the energy range of PANDA is the large production
cross section of hyperon-antihyperon pairs \cite{panda_phys}. At its
full luminosity the production rate of $\Pay\Py$-pairs range
from a few 100 per second for the $\PagX\PgX$-channel, up to a
few thousand per second for the $\PagL\PgL$-channel in the elementary
$\Pap\Pp$-reaction. Due to the strong absorption of
antibaryons in nuclei this production rate will be lowered
depending on the size of the target nucleus in antiproton-nucleus
collisions. According to the GiBUU calculations
presented above, for a typical medium size target nucleus
like $^{20}$Ne still several hundreds free $\PagL\PgL$-pairs can be
produced per second. For a nuclear target in this mass
range and at maximum interaction rate, approximately
10 reconstructed $\PagL\PgL$-pairs per second are expected \cite{panda_tech}.
Therefore, a measurement period of about one hour will
provide a statistics exceeding that of the GiBUU simulations
shown above. This will be sufficient to reach a
precision of about 10\% for the scaling factor $\xi_{\PagL}$ of the antilambda
potential. These numbers illustrate that even
on rather pessimistic assumption about the luminosity
and/or the availability of the antiproton beam during the
commissioning phase of the PANDA experiment, one can
reach unique and relevant information on the behavior of
strange antibaryons in nuclei shortly after the delivery
of the first antiproton beam at FAIR. 

This work was supported in part by European Community Research Infrastructure Integrating Activity
'Study of Strongly Interacting Matter' HadronPhysics3 (SPHERE) under the FP7.

%\section*{References}

\end{document}